



\font\bigbf=cmbx10 scaled\magstep2

\font\twelverm=cmr10 scaled 1200    \font\twelvei=cmmi10 scaled 1200
\font\twelvesy=cmsy10 scaled 1200   \font\twelveex=cmex10 scaled 1200
\font\twelvebf=cmbx10 scaled 1200   \font\twelvesl=cmsl10 scaled 1200
\font\twelvett=cmtt10 scaled 1200   \font\twelveit=cmti10 scaled 1200

\skewchar\twelvei='177   \skewchar\twelvesy='60


\def\twelvepoint{\normalbaselineskip=12.4pt
  \abovedisplayskip 12.4pt plus 3pt minus 9pt
  \belowdisplayskip 12.4pt plus 3pt minus 9pt
  \abovedisplayshortskip 0pt plus 3pt
  \belowdisplayshortskip 7.2pt plus 3pt minus 4pt
  \smallskipamount=3.6pt plus1.2pt minus1.2pt
  \medskipamount=7.2pt plus2.4pt minus2.4pt
  \bigskipamount=14.4pt plus4.8pt minus4.8pt
  \def\rm{\fam0\twelverm}          \def\it{\fam\itfam\twelveit}%
  \def\sl{\fam\slfam\twelvesl}     \def\bf{\fam\bffam\twelvebf}%
  \def\mit{\fam 1}                 \def\cal{\fam 2}%
  \def\tt{\twelvett}
  \textfont0=\twelverm   \scriptfont0=\tenrm   \scriptscriptfont0=\sevenrm
  \textfont1=\twelvei    \scriptfont1=\teni    \scriptscriptfont1=\seveni
  \textfont2=\twelvesy   \scriptfont2=\tensy   \scriptscriptfont2=\sevensy
  \textfont3=\twelveex   \scriptfont3=\twelveex 
 \scriptscriptfont3=\twelveex
  \textfont\itfam=\twelveit
  \textfont\slfam=\twelvesl
  \textfont\bffam=\twelvebf \scriptfont\bffam=\tenbf
  \scriptscriptfont\bffam=\sevenbf
  \normalbaselines\rm}



\def\beginlinemode{\endmode
  \begingroup\parskip=0pt 
\obeylines\def\\{\par}\def\endmode{\par\endgroup}}
\def\beginparmode{\endmode
  \begingroup \def\endmode{\par\endgroup}}
\let\endmode=\par
{\obeylines\gdef\
{}}
\def\singlespace{\baselineskip=\normalbaselineskip}
\def\oneandathirdspace{\baselineskip=\normalbaselineskip
  \multiply\baselineskip by 4 \divide\baselineskip by 3}

\def\doublespace{\baselineskip=
\normalbaselineskip \multiply\baselineskip by 2}

\newcount\firstpageno
\firstpageno=1
\footline={\ifnum\pageno<\firstpageno{\hfil}%
\else{\hfil\twelverm\folio\hfil}\fi}
\let\rawfootnote=\footnote              
\def\footnote#1#2{{\rm\singlespace\parindent=0pt\rawfootnote{#1}{#2}}}
\def\raggedcenter{\leftskip=4em plus 12em \rightskip=\leftskip
  \parindent=0pt \parfillskip=0pt \spaceskip=.3333em \xspaceskip=.5em
  \pretolerance=9999 \tolerance=9999
  \hyphenpenalty=9999 \exhyphenpenalty=9999 }
\def\dateline{\rightline{\ifcase\month\or
  January\or February\or March\or April\or May\or June\or
  July\or August\or September\or October\or November\or December\fi
  \space\number\year}}
\def\received{\vskip 3pt plus 0.2fill
 \centerline{\sl (Received\space\ifcase\month\or
  January\or February\or March\or April\or May\or June\or
  July\or August\or September\or October\or November\or December\fi
  \qquad, \number\year)}}


\hsize=6.5truein
\vsize=8.9truein
\voffset=0.0truein
\parskip=\medskipamount
\twelvepoint            
\oneandathirdspace           
\overfullrule=0pt       



\def\title                      
  {\null\vskip 3pt plus 0.2fill
   \beginlinemode \doublespace \raggedcenter \bigbf}

\def\author                     
  {\vskip 3pt plus 0.2fill \beginlinemode
   \singlespace \raggedcenter}

\def\affil                      
  {\vskip 4pt 
\beginlinemode
   \singlespace \raggedcenter \sl}

\def\abstract                   
  {\vskip 3pt plus 0.3fill \beginparmode
   \oneandathirdspace\narrower}

\def\endtitlepage               
  {\endpage                     
   \body}

\def\body                       
  {\beginparmode}               

\def\head#1{                    
  \vskip 0.25truein     
 {\immediate\write16{#1}
   \noindent{\bf{#1}}\par}
   \nobreak\vskip 0.125truein\nobreak}

\def\subhead#1{                 
  \vskip 0.25truein             
  \noindent{{\it {#1}} \par}
   \nobreak\vskip 0.15truein\nobreak}

\def\refto#1{[#1]}           

\def\references                 
  {\subhead{\bf References}         
   \beginparmode
   \frenchspacing \parindent=0pt \leftskip=1truecm
   \oneandathirdspace\parskip=8pt plus 3pt
 \everypar{\hangindent=\parindent}}

\gdef\refis#1{\indent\hbox to 0pt{\hss#1.~}}    

\gdef\journal#1, #2, #3, #4#5#6#7{               
    {\sl #1~}{\bf #2}, #3 (#4#5#6#7)}           

\def\refstylenp{                
  \gdef\refto##1{ [##1]}                                
  \gdef\refis##1{\indent\hbox to 0pt{\hss##1)~}}        
  \gdef\journal##1, ##2, ##3, ##4 {                     
     {\sl ##1~}{\bf ##2~}(##3) ##4 }}

\def\refstyleprnp{              
  \gdef\refto##1{ [##1]}                                
  \gdef\refis##1{\indent\hbox to 0pt{\hss##1)~}}        
  \gdef\journal##1, ##2, ##3, 1##4##5##6{               
    {\sl ##1~}{\bf ##2~}(1##4##5##6) ##3}}

\def\prd{\journal Phys. Rev. D, }

\def\prl{\journal Phys. Rev. Lett., }

\def\jmp{\journal J. Math. Phys., }

\def\npb{\journal Nucl. Phys. B, }

\def\cqg{\journal Class. Quantum Grav., }

\def\ann{\journal Ann. Phys., }

\def\endreferences{\body}

\def\figurecaptions             
  { \beginparmode
   \subhead{Figure Captions}
}

\def\endpage                    
  {\vfill\eject}

\def\endpaper                   
  {\endmode\vfill\supereject}

\def\endit
  {\endpaper\end}

\def\hook{\mathbin{\raise2.5pt\hbox{\hbox{{\vbox{\hrule height.4pt 
width6pt depth0pt}}}\vrule height3pt width.4pt depth0pt}\,}}
\def\today{\number\day\ \ifcase\month\or
     January\or February\or March\or April\or May\or June\or
     July\or August\or September\or October\or November\or
     December\space \fi\ \number\year}
\def\date{\noindent{\tt 
     Date typeset: \today\par\bigskip}}
\def\ref#1{Ref. #1}                     
\def\Ref#1{Ref. #1}                     

\def\frac#1#2{{\textstyle{#1 \over #2}}}
\def\half{{\textstyle{ 1\over 2}}}
\def\>{\rangle}
\def\<{\langle}
\def\eg{{\it e.g.,\ }}

\def\ie{{\it i.e.,\ }}

\def\etc{{\it etc.}}

\def\sla{\raise.15ex\hbox{$/$}\kern-.57em}
\def\leaderfill{\leaders\hbox to 1em{\hss.\hss}\hfill}
\def\twiddle{\lower.9ex\rlap{$\kern-.1em\scriptstyle\sim$}}
\def\bigtwiddle{\lower1.ex\rlap{$\sim$}}
\def\gtwid{
\mathrel{\raise.3ex\hbox{$>$\kern-.75em\lower1ex\hbox{$\sim$}}}}
\def\ltwid{\mathrel{\raise.3ex\hbox
{$<$\kern-.75em\lower1ex\hbox{$\sim$}}}}
\def\square{\kern1pt\vbox{\hrule height 1.2pt\hbox
{\vrule width 1.2pt\hskip 3pt
   \vbox{\vskip 6pt}\hskip 3pt\vrule width 0.6pt}
\hrule height 0.6pt}\kern1pt}

\def\m@th{\mathsurround=0pt }
\def\leftrightarrowfill{$\m@th \mathord\leftarrow \mkern-6mu
 \cleaders\hbox{$\mkern-2mu \mathord- \mkern-2mu$}\hfill
 \mkern-6mu \mathord\rightarrow$}
\def\overleftrightarrow#1{\vbox{\ialign{##\crcr
     \leftrightarrowfill\crcr\noalign{\kern-1pt\nointerlineskip}
     $\hfil\displaystyle{#1}\hfil$\crcr}}}


\font\titlefont=cmr10 scaled\magstep3

\def\martinstyletitle                      
  {\null\vskip 3pt plus 0.2fill
   \beginlinemode \doublespace \raggedcenter \titlefont}

\font\twelvesc=cmcsc10 scaled 1200

\def\author                     
  {\vskip 3pt plus 0.2fill \beginlinemode
   \singlespace \raggedcenter\twelvesc}


\def\endtitle{\body}
\def\endauthor{\body}
\def\endaffil{\body}

\def\heading                            
  {\vskip 0.5truein plus 0.1truein      
\endheading
   \beginparmode \def\\{\par} \parskip=0pt \singlespace \raggedcenter}

\def\endheading
  {\par\nobreak\vskip 0.25truein\nobreak\beginparmode}

\def\subheading                         
  {\vskip 0.25truein plus 0.1truein     
   \beginlinemode \singlespace \parskip=0pt \def\\{\par}\raggedcenter}

\def\tag#1$${\eqno(#1)$$}

\def\align#1$${\eqalign{#1}$$}

\def\aligntag#1$${\gdef\tag##1\\{&(##1)\cr}\eqalignno{#1\\}$$
  \gdef\tag##1$${\eqno(##1)$$}}

\def\endaligntag{}

\def\overset #1\to#2{{\mathop{#2}\limits^{#1}}}
\def\underset#1\to#2{{\let\next=#1\mathpalette\undersetpalette#2}}
\def\undersetpalette#1#2{\vtop{\baselineskip0pt
\ialign{$\mathsurround=0pt #1\hfil##\hfil$\crcr#2\crcr\next\crcr}}}


\def\ref#1{Ref.~#1}                     
\def\Ref#1{Ref.~#1}                     
\def\[#1]{[\cite{#1}]}
\def\cite#1{{#1}}
\def\(#1){(\call{#1})}
\def\call#1{{#1}}
\def\taghead#1{}
\def\frac#1#2{{#1 \over #2}}
\def\half{{\frac 12}}

\def\12{{1\over2}}
\def\eg{{\it e.g.,\ }}

\def\ie{{\it i.e.,\ }}

\def\etc{{\it etc.\ }}

\def\cf{{\sl cf.\ }}
\def\sla{\raise.15ex\hbox{$/$}\kern-.57em}
\def\leaderfill{\leaders\hbox to 1em{\hss.\hss}\hfill}
\def\twiddle{\lower.9ex\rlap{$\kern-.1em\scriptstyle\sim$}}
\def\bigtwiddle{\lower1.ex\rlap{$\sim$}}
\def\gtwid{\mathrel{\raise.3ex\hbox{$>$
\kern-.75em\lower1ex\hbox{$\sim$}}}}
\def\ltwid{\mathrel{\raise.3ex\hbox{$<$
\kern-.75em\lower1ex\hbox{$\sim$}}}}
\def\square{\kern1pt\vbox{\hrule height 1.2pt\hbox
{\vrule width 1.2pt\hskip 3pt
   \vbox{\vskip 6pt}\hskip 3pt\vrule width 0.6pt}
\hrule height 0.6pt}\kern1pt}
\def\tdot#1{\mathord{\mathop{#1}\limits^{\kern2pt\ldots}}}

\def\pmb#1{\setbox0=\hbox{#1}%
  \kern-.025em\copy0\kern-\wd0
  \kern  .05em\copy0\kern-\wd0
  \kern-.025em\raise.0433em\box0 }

\catcode`@=11
\newcount\tagnumber\tagnumber=0

\immediate\newwrite\eqnfile
\newif\if@qnfile\@qnfilefalse
\def\write@qn#1{}
\def\writenew@qn#1{}
\def\w@rnwrite#1{\write@qn{#1}\message{#1}}
\def\@rrwrite#1{\write@qn{#1}\errmessage{#1}}

\def\taghead#1{\gdef\t@ghead{#1}\global\tagnumber=0}
\def\t@ghead{}

\expandafter\def\csname @qnnum-3\endcsname
  {{\t@ghead\advance\tagnumber by -3\relax\number\tagnumber}}
\expandafter\def\csname @qnnum-2\endcsname
  {{\t@ghead\advance\tagnumber by -2\relax\number\tagnumber}}
\expandafter\def\csname @qnnum-1\endcsname
  {{\t@ghead\advance\tagnumber by -1\relax\number\tagnumber}}
\expandafter\def\csname @qnnum0\endcsname
  {\t@ghead\number\tagnumber}
\expandafter\def\csname @qnnum+1\endcsname
  {{\t@ghead\advance\tagnumber by 1\relax\number\tagnumber}}
\expandafter\def\csname @qnnum+2\endcsname
  {{\t@ghead\advance\tagnumber by 2\relax\number\tagnumber}}
\expandafter\def\csname @qnnum+3\endcsname
  {{\t@ghead\advance\tagnumber by 3\relax\number\tagnumber}}

\def\equationfile{%
  \@qnfiletrue\immediate\openout\eqnfile=\jobname.eqn%
  \def\write@qn##1{\if@qnfile\immediate\write\eqnfile{##1}\fi}
  \def\writenew@qn##1{\if@qnfile\immediate\write\eqnfile
    {\noexpand\tag{##1} = (\t@ghead\number\tagnumber)}\fi}
}

\def\callall#1{\xdef#1##1{#1{\noexpand\call{##1}}}}
\def\call#1{\each@rg\callr@nge{#1}}

\def\each@rg#1#2{{\let\thecsname=#1\expandafter\first@rg#2,\end,}}
\def\first@rg#1,{\thecsname{#1}\apply@rg}
\def\apply@rg#1,{\ifx\end#1\let\next=\relax%
\else,\thecsname{#1}\let\next=\apply@rg\fi\next}

\def\callr@nge#1{\calldor@nge#1-\end-}
\def\callr@ngeat#1\end-{#1}
\def\calldor@nge#1-#2-{\ifx\end#2\@qneatspace#1 %
  \else\calll@@p{#1}{#2}\callr@ngeat\fi}
\def\calll@@p#1#2{\ifnum#1>#2{\@rrwrite
{Equation range #1-#2\space is bad.}
\errhelp{If you call a series of equations by the notation M-N, then M and
N must be integers, and N must be greater than or equal to M.}}\else %
{\count0=#1\count1=
#2\advance\count1 by1\relax\expandafter\@qncall\the\count0,%
  \loop\advance\count0 by1\relax%
    \ifnum\count0<\count1,\expandafter\@qncall\the\count0,%
  \repeat}\fi}

\def\@qneatspace#1#2 {\@qncall#1#2,}
\def\@qncall#1,{\ifunc@lled{#1}{\def\next{#1}\ifx\next\empty\else
  \w@rnwrite{Equation number \noexpand\(>>#1<<) 
has not been defined yet.}
  >>#1<<\fi}\else\csname @qnnum#1\endcsname\fi}

\let\eqnono=\eqno
\def\eqno(#1){\tag#1}
\def\tag#1$${\eqnono(\displayt@g#1 )$$}

\def\aligntag#1\endaligntag
  $${\gdef\tag##1\\{&(##1 )\cr}\eqalignno{#1\\}$$
  \gdef\tag##1$${\eqnono(\displayt@g##1 )$$}}

\def\eqalignno#1{\displ@y \tabskip\centering
  \halign to\displaywidth{\hfil$\displaystyle{##}$\tabskip\z@skip
    &$\displaystyle{{}##}$\hfil\tabskip\centering
    &\llap{$\displayt@gpar##$}\tabskip\z@skip\crcr
    #1\crcr}}

\def\displayt@gpar(#1){(\displayt@g#1 )}

\def\displayt@g#1 {\rm\ifunc@lled{#1}\global\advance\tagnumber by1
        {\def\next{#1}\ifx\next\empty\else\expandafter
        \xdef\csname
 @qnnum#1\endcsname{\t@ghead\number\tagnumber}\fi}%
  \writenew@qn{#1}\t@ghead\number\tagnumber\else
        {\edef\next{\t@ghead\number\tagnumber}%
        \expandafter\ifx\csname @qnnum#1\endcsname\next\else
        \w@rnwrite{Equation \noexpand\tag{#1} is 
a duplicate number.}\fi}%
  \csname @qnnum#1\endcsname\fi}

\def\ifunc@lled#1{\expandafter\ifx\csname @qnnum#1\endcsname\relax}

\let\@qnend=\end\gdef\end{\if@qnfile
\immediate\write16{Equation numbers 
written on []\jobname.EQN.}\fi\@qnend}

\catcode`@=12

\catcode`@=11
\newcount\r@fcount \r@fcount=0
\newcount\r@fcurr
\immediate\newwrite\reffile
\newif\ifr@ffile\r@ffilefalse
\def\w@rnwrite#1{\ifr@ffile\immediate\write\reffile{#1}\fi\message{#1}}

\def\writer@f#1>>{}
\def\referencefile{
  \r@ffiletrue\immediate\openout\reffile=\jobname.ref%
  \def\writer@f##1>>{\ifr@ffile\immediate\write\reffile%
    {\noexpand\refis{##1} = \csname r@fnum##1\endcsname = %
     \expandafter\expandafter\expandafter\strip@t\expandafter%
     \meaning\csname r@ftext
\csname r@fnum##1\endcsname\endcsname}\fi}%
  \def\strip@t##1>>{}}

\def\citeall#1{\xdef#1##1{#1{\noexpand\cite{##1}}}}
\def\cite#1{\each@rg\citer@nge{#1}}	

\def\each@rg#1#2{{\let\thecsname=#1\expandafter\first@rg#2,\end,}}
\def\first@rg#1,{\thecsname{#1}\apply@rg}	
\def\apply@rg#1,{\ifx\end#1\let\next=\relax
\else,\thecsname{#1}\let\next=\apply@rg\fi\next}

\def\citer@nge#1{\citedor@nge#1-\end-}	
\def\citer@ngeat#1\end-{#1}
\def\citedor@nge#1-#2-{\ifx\end#2\r@featspace#1 
  \else\citel@@p{#1}{#2}\citer@ngeat\fi}	
\def\citel@@p#1#2{\ifnum#1>#2{\errmessage{Reference range #1-
#2\space is bad.}%
    \errhelp{If you cite a series of references by the notation M-N, then M 
and
    N must be integers, and N must be greater than or equal to M.}}\else%
 {\count0=#1\count1=#2\advance\count1 
by1\relax\expandafter\r@fcite\the\count0,
  \loop\advance\count0 by1\relax
    \ifnum\count0<\count1,\expandafter\r@fcite\the\count0,%
  \repeat}\fi}

\def\r@featspace#1#2 {\r@fcite#1#2,}	
\def\r@fcite#1,{\ifuncit@d{#1}
    \newr@f{#1}%
    \expandafter\gdef\csname r@ftext\number\r@fcount\endcsname%
                     {\message{Reference #1 to be supplied.}%
                      \writer@f#1>>#1 to be supplied.\par}%
 \fi%
 \csname r@fnum#1\endcsname}
\def\ifuncit@d#1{\expandafter\ifx\csname r@fnum#1\endcsname\relax}%
\def\newr@f#1{\global\advance\r@fcount by1%
    \expandafter\xdef\csname r@fnum#1\endcsname{\number\r@fcount}}

\let\r@fis=\refis			
\def\refis#1#2#3\par{\ifuncit@d{#1}
   \newr@f{#1}%
   \w@rnwrite{Reference #1=\number\r@fcount\space is not cited up to
 now.}\fi%
  \expandafter
\gdef\csname r@ftext\csname r@fnum#1\endcsname\endcsname%
  {\writer@f#1>>#2#3\par}}

\def\ignoreuncited{
   \def\refis##1##2##3\par{\ifuncit@d{##1}%
    \else\expandafter\gdef
\csname r@ftext\csname r@fnum##1\endcsname\endcsname%
     {\writer@f##1>>##2##3\par}\fi}}

\def\r@ferr{\endreferences\errmessage{I was expecting to see
\noexpand\endreferences before now;  I have inserted it here.}}
\let\r@ferences=\references
\def\references{\r@ferences\def\endmode{\r@ferr\par\endgroup}}

\let\endr@ferences=\endreferences
\def\endreferences{\r@fcurr=0
  {\loop\ifnum\r@fcurr<\r@fcount
    \advance\r@fcurr by 
1\relax\expandafter\r@fis\expandafter{\number\r@fcurr}%
    \csname r@ftext\number\r@fcurr\endcsname%
  \repeat}\gdef\r@ferr{}\endr@ferences}


\let\r@fend=\endpaper\gdef\endpaper{\ifr@ffile
\immediate\write16{Cross References written on 
[]\jobname.REF.}\fi\r@fend}

\catcode`@=12

\citeall\refto		
\citeall\ref		%
\citeall\Ref		%

\ignoreuncited

\def\Hphys{{\cal H}_{phys}}

\pageno=0
\line{\hfill June 2002}
\title
Quantum Dynamics 
of the 
Polarized Gowdy ${\bf T}^3$ Model
\endtitle
\author
C. G. Torre
\endauthor
\affil
Department of Physics
Utah State University
Logan, UT 84322-4415 USA
torre@cc.usu.edu
\endaffil

\abstract
The polarized Gowdy ${\bf T}^3$ vacuum spacetimes are characterized, modulo gauge, by a ``point particle'' degree of freedom  and a function $\varphi$ that satisfies a linear field equation and a non-linear constraint. The  quantum Gowdy model has been defined  by using a representation for $\varphi$ on a Fock space $\cal F$. Using this quantum model, it has recently been shown that the dynamical evolution determined by the linear field equation for $\varphi$ is not unitarily implemented on $\cal F$. In this paper: (1) We derive the classical and quantum model using the ``covariant phase space'' formalism. (2) We show that time evolution is not unitarily implemented even on the physical Hilbert space of states ${\cal H} \subset {\cal F}$ defined by the quantum constraint. (3)  We show that the spatially smeared canonical coordinates and momenta as well as the time-dependent Hamiltonian for $\varphi$ are well-defined, self-adjoint operators for all time, admitting the usual probability interpretation despite the lack of unitary dynamics.

\endtitlepage

\taghead{1.}
\head{1. Introduction}

Over the past 30 years, spacetimes admitting two commuting Killing vector fields have been studied repeatedly as ``midi-superspace'' models for canonical quantum gravity; see, for example,  references [1--11].\setbox0=\hbox{\refto{Kuchar1971,Misner1973,Berger,Husain1987,Husain1989,Neville1993,CGT1996,Ashtekar1996b,Korotkin1998,Pierri,Corichi2002}} 
These models admit an infinite number of degrees of freedom --- they are field theories --- and as such they are more sophisticated than the ``mini-superspace'' models, which are mechanical models with a finite number of degrees of freedom.  In particular, the quantized midi-superspace models can bring into play intrinsically quantum field theoretic features which have no analogs in quantum mechanics. One of these features -- the failure of time evolution to be unitarily implemented -- is the impetus for the present paper. 

Of the midi-superspace models the Gowdy class \refto{Gowdy1974} is interesting since it defines an inhomogeneous cosmology including a big bang or a big crunch.  It was first studied as a model of quantum gravity by Misner \refto{Misner1973} and  Berger \refto{Berger}, who explored a variety of approaches to defining the quantum theory and extracting physics from the polarized Gowdy ${\bf T}^3$ model.  This model arises by assuming spacetime is not flat, that it has the topology ${\bf R}\times {\bf  T}^3$, and that there is an Abelian 2 parameter isometry group with spacelike  orbits ${\bf T}^2\subset {\bf T}^3$ generated by a pair of commuting, hypersurface-orthogonal Killing vector fields.  In this setting, the vacuum Einstein equations imply that, modulo gauge and a ``point particle'' degree of freedom, the classical dynamics of this model is governed by a function $\varphi$ that satisfies a linear field equation and a non-linear constraint. The field equation is equivalent to that of a one-dimensional symmetry reduction of a massless free scalar field propagating on a flat spacetime $(M,g)$. The spacetime $(M,g)$ can be viewed as the causal region of (a compactification of) a three-dimensional version of Misner spacetime.\footnote*{For a discussion of Misner spacetime, see \refto{Hawking1973}.}  The Gowdy time foliation equips the $(M,g)$ with a foliation by  expanding (or contracting) spatial sections, along which the time evolution of $\varphi$ is given by the linear field equation. The constraint requires the total momentum of the scalar field $\varphi$ to vanish.    One of the quantizations of $\varphi$ studied by Berger, and subsequently studied by Husain \refto{Husain1987},  Pierri \refto{Pierri}, and Corichi {\it et al} \refto{Corichi2002}, is equivalent to the restriction of  the standard quantum theory of a massless free scalar field on three-dimensional Minkowski space to the compactified Misner spacetime $(M,g)$. This defines a Fock space representation for $\varphi$. The total field momentum can be defined in this representation and physical states are eigenstates of it with zero eigenvalue. Thus the polarized Gowdy ${\bf T}^3$ model  is defined as a constrained quantum field theory. 

In reference \refto{Corichi2002}, it was observed that the time evolution of $\varphi$, as defined by its linear field equation, cannot be implemented as a unitary transformation on the Fock space described above.  This sort of phenomenon, which is not unexpected in quantum field theory \refto{Segal1963, Helfer1999}, has been seen in other, related settings \refto{Helfer1996, CGT1999c}. The lack of unitary dynamics leads the authors of \refto{Corichi2002} to conclude that the quantized model under consideration is not physically viable.  
In this paper we will extend the results of \refto{Corichi2002} in 3 significant ways, which will be described in the following paragraphs.

First, the classical analysis we develop in \S 2  as the underpinning for the  quantum theory (\S 3) is quite different from the standard Dirac techniques utilized in \refto{Pierri, Corichi2002} since we formulate the model  using the   ``covariant phase space'' formalism \refto{covps}.  This feature of our work is perhaps of some intrinsic interest since it represents a non-trivial application of that formalism. But the principal utility of the covariant phase space approach to the Gowdy model is that it allows for an independent, relatively simple construction of the  ``deparametrized'' dynamical system, which is not entirely straightforward in the Dirac-type of  approach featuring in \refto{Pierri, Corichi2002} owing to the presence of the point particle degrees of freedom which are subsequently mixed in with the time variable during the deparametrization process.  The covariant phase space formalism allows one to work directly with the classical spacetime and this makes it very easy to keep track of the field degrees of freedom, the point particle degrees of freedom and the choice of time that is used in the model.  

Second, we show in \S 4 that the proof of the failure of time evolution to be unitarily implemented on the initial, ``auxiliary'' Fock space can be extended to apply to the physical Hilbert space of states defined by the momentum constraint.  {\it A priori}, it is possible that the dynamical evolution is unitarily implemented on the physical Hilbert space but not on the auxiliary Hilbert space.  In this scenario the polarized Gowdy ${\bf T}^3$ model would have unitary dynamics, physically speaking, with the non-unitarity found in \refto{Corichi2002} just being a technical complication.  We show, however, that this is not the case.  

Third, we show in \S 5-6 that, despite the failure of time evolution to be unitarily implemented, a number of basic operators have entirely satisfactory behavior.  In particular, the canonical coordinates and momentum of the quantum field $\varphi$, and all their derivatives (when spatially smeared with smooth functions) are at each time well-defined self-adjoint operators with continuous spectrum on the real line.  Therefore, if not for the existence of the momentum constraint, which restricts the class of physical operators, the canonical field  operators would represent observables with a perfectly acceptable physical interpretation in the Heisenberg picture.  Operators representing physical observables can be obtained from the canonical field operators by projection into the physical Hilbert space \refto{Pierri}, and the self-adjointness of the field operators implies that these operators are again physically acceptable, despite the lack of unitary dynamics on the physical Hilbert space.  Moreover, we show that the one-parameter family of Hamiltonians for the deparametrized Gowdy model can be defined as self-adjoint operators both on the auxiliary Fock space and on the physical Hilbert space, despite the fact that the dynamical evolution they generate is not unitarily implementable. 

The results summarized in the previous paragraph indicate that it may be possible to considerably soften the conclusions reached in \refto{Corichi2002} concerning the physical viability of the quantum Gowdy model.  We discuss the situation in \S 7.

\taghead{2.}
\head{2. Polarized Gowdy model: classical theory}

Fix coordinates $(t,x,y,z)$ on $M={\bf R}^+\times {\bf T}^3$, where $t>0$ and $0< x,y,z<2\pi$. The polarized Gowdy ${\bf T}^3$ metrics are defined by \refto{Gowdy1974, Chrusciel1990}
$$
g = l\left\{e^{(\gamma-\varphi)}(-dt\otimes dt + dx\otimes dx) + t^2 e^{-\varphi}dy\otimes dy + e^\varphi dz\otimes dz\right\},
\tag Gowdy_metric
$$
where $l>0$ is a constant and both $\gamma=\gamma(t,x)$ and $\varphi=\varphi(t,x)$ are periodic functions of $x$ with period $2\pi$. The space of Gowdy metrics is thus parametrized by $(l,\gamma,\varphi)$. 

The vacuum Einstein equations, when restricted to \(Gowdy_metric), are equivalent to
$$
-\varphi_{,tt} - {1\over t}\varphi_{,t} + \varphi_{,xx} = 0,
\tag phi_eq
$$
$$
\gamma_{,t} = {t\over 2}\left(\varphi_{,t}^2 + \varphi_{,x}^2\right)\equiv {\cal E},\quad
\gamma_{,x} = t\varphi_{,t}\varphi_{,x}\equiv \Pi.
\tag gamma_eq
$$
All smooth solutions to $\(phi_eq)$ are of the form
$$
\varphi(t,x)={1\over \sqrt{2\pi l}}(q + p \ln t)
+{1\over2\sqrt{2 l}} \sum_{n=-\infty\atop n\neq 0}^\infty\left( a_n  H_0(|n|t)e^{in x} + a^*_n  H^*_0(|n|t)e^{-in x}\right),
\tag psi_soln
$$
where $H_0$ is a Hankel function of the second kind\footnote*{We could equally well use Hankel functions of the first kind, which is the choice made in \refto{Pierri, Corichi2002}. The choice used here has the feature that its notion of positive frequency agrees with that defined by the timelike Killing vector field of Minkowski space when the field $\varphi$ is interpreted as a function on the interior of the future light cone. This is the same convention used in the ``Schmidt model'' \refto{Beetle1998}.  No results in this paper depend upon which type of Hankel function is used.} and the sequence $\{a_n\}$ is rapidly decreasing, \ie its elements approach zero faster than the reciprocal of any polynomial in $n$ as $n\to\pm \infty$. Given \(psi_soln), the solutions to \(gamma_eq) can be expressed in the form
$$
\gamma(t,x) = \gamma_0 + \int_{t_0}^t dt^\prime {\cal E}( t^\prime,x) + \int_{x_0}^{x} dx^\prime \Pi(t_0,x^\prime),
\tag gamma_soln
$$
where $\gamma_0$ is a constant.  The metric function $\gamma$ is thus parametrized by $\gamma_0$ and $\varphi$. (The quantities  $(t_0,x_0)$ can be fixed arbitrarily; different choices of them merely redefine  $\gamma_0$.)     $\gamma$ is periodic in $x$ if and only if  
$$
{\cal C}\equiv l\int_0^{2\pi} dx\, \Pi(x) = \sum_{n=-\infty}^\infty n |a_n|^2= 0,
\tag constraint
$$
which can be viewed as the sole remnant of the Hamiltonian and momentum constraints. Note that ${\cal C}$ is independent of $t$, so that the constraint \(constraint) need only be imposed at a single value of $t$.

According to \(psi_soln)--\(constraint), the set of smooth solutions  to the Einstein equations  for the polarized Gowdy metrics  can be identified with the set 
$$\Gamma=\{(l,\gamma_0,q,p,a_n,a_n^*),\ n=\pm 1,\pm2,\ldots|{\cal C}=0\}.
\tag Gamma
$$ 
 The set $\Gamma$ has a pre-symplectic structure naturally defined by the Einstein-Hilbert  action as follows.\footnote*{Using the results of \refto{CGT2002} it can be shown that the polarized Gowdy isometry group satisfies the ``Principle of Symmetric Criticality''. This implies that any generally covariant local variational principle for a spacetime metric, when restricted to the set of all metrics admitting the polarized Gowdy isometry group, yields the correct set of reduced field equations.}
With a convenient normalization,  the Einstein-Hilbert action takes the following, remarkably simple form when restricted to the Gowdy metrics \(Gowdy_metric):
$$
\eqalign{
S[l,\gamma,\varphi]&=  \left({1\over 2\pi}\right)^2 \int_{M}\sqrt{g} R\cr  
&=-l\int_{t_1}^{t_2}dt\, \int_0^{2\pi}dx\, t\left\{
-\gamma_{,tt}  +\varphi_{,tt} +{1\over t} \varphi_{,t} -\half \varphi_{,t}^2 + \half \varphi_{,x}^2
\right\}.}
\tag action
$$
Of course, one cannot obtain all the field equations \(phi_eq)--\(gamma_eq) as critical points of \(action) since gauge fixing conditions have been used  to define \(Gowdy_metric).\footnote\dag{One can get only the $\varphi$ field equations from varying $\varphi$ in \(action). The equations obtained from varying $\gamma$ are trivial since $\gamma$ only appears as a divergence. The equation coming from varying $l$ is non-trivial, but is automatically satisfied when $\varphi$ and $\gamma$ satisfy the equations \(phi_eq)-\(gamma_eq).}  Nevertheless, as we shall see, this action is suitable for defining a pre-symplectic structure on  $\Gamma$.\footnote\ddag{From the point of view of a traditional  Dirac type of Hamiltonian analysis, one can interpret this action and its pre-symplectic structure as corresponding to the result of a  partial gauge fixing and deparametrization, with all of the first class constraints except \(constraint) being rendered second class, followed by the use of Dirac brackets.} 
%
%
 Following the general prescription of \refto{covps},  vary the action \(action) and extract the resulting boundary term, which defines a ($t$-dependent) 1-form on the space of dynamical variables $(l,\gamma,\varphi)$.  Pull back this 1-form to $\Gamma$ to get the pre-symplectic potential. At a point $(l,\gamma_0,\varphi)\in \Gamma$, the value of this 1-form on a tangent vector $(\delta l,\delta\gamma_0,\delta \varphi)$ is given by
$$
\eqalign{
\theta(\delta l,\delta\gamma_0,\delta \varphi)
&=\int_0^{2\pi}dx\, l\left(t\delta\gamma_{,t} - t\delta\varphi_{,t} - \delta\gamma + t\varphi_{,t}\delta\varphi
\right)
\cr
&=\int_0^{2\pi}dx\, l\left\{[-\delta\gamma_0 + t\delta{\cal E} - t\delta\varphi_{,t}
-\int_{x_0}^x dx^\prime\, \delta \Pi(t_0,x^\prime) - \int_{t_0}^{t} dt^\prime \delta{\cal E}(t^\prime,x)
+ t\varphi_{,t}\delta\varphi\right\}\cr
&=l\Bigg(-2\pi\delta\gamma_0 + t \delta E(t) - \sqrt{2\pi} \delta \left({p\over\sqrt{l}}\right)  -\int_0^{2\pi}dx\int_{x_0}^x dx^\prime\delta \Pi(t_0,x^\prime)-\int_{t_0}^{t} dt^\prime \delta E(t^\prime)\cr
&\quad+ \int_0^{2\pi}dx\,t\varphi_{,t}\delta\varphi
\Bigg)\cr
&= l \delta\left(-2\pi\gamma_0 + t  E(t) - \sqrt{2\pi\over l}  p  
-\int_0^{2\pi}dx\int_{x_0}^x dx^\prime\,  \Pi(t_0,x^\prime)-\int_{t_0}^{t} dt^\prime  E(t^\prime)\right)\cr
&\quad+ \int_0^{2\pi}dx\,l\,t\varphi_{,t}\delta\varphi}
\tag theta
$$
where
$$
E(t) = \int_0^{2\pi} dx\, {\cal E}(t,x).
\tag Eoft
$$
One point to keep in mind here is that the fields $\varphi$ and their variations $\delta\varphi$ are subject to the field equations and constraints.
The pre-symplectic two-form is obtained from the symplectic potential by exterior differentiation. At the point $(l,\gamma_0,\varphi)\in \Gamma$, the two-form evaluated on a pair of vectors, $(\delta l, \delta\gamma_0,\delta\varphi)$ and $(\tilde\delta l,\tilde\delta\gamma_0,\tilde\delta\varphi)$, is given by
$$
\omega(\delta l, \delta\gamma_0,\delta\varphi; \tilde\delta l,\tilde\delta\gamma_0,\tilde\delta\varphi)
= \delta l \tilde \delta \xi  - \tilde\delta l \delta \xi  + \int_0^{2\pi} dx\, \left(\delta P_\phi \tilde \delta\phi - \tilde \delta P_\phi \delta \phi\right),
\tag symplectic
$$
where
$$
\phi:= \sqrt{l} \varphi ,\quad P_\phi:=t\sqrt{l}\varphi_{,t},
\tag can_field
$$
and
$$
\eqalign{
\xi:&=-2\pi\gamma_0 + t  E(t) - \sqrt{2\pi\over l}  p  
-\int_0^{2\pi}dx\int_{x_0}^x dx^\prime\,  \Pi(t_0,x^\prime)-\int_{t_0}^{t} dt^\prime  E(t^\prime)\cr
 &\quad+ \half {1\over l}\int_0^{2\pi}dx\, P_\phi(t,x) \phi(t,x).}
\tag xi
$$
From the field equations  it follows that $\xi$ (like $l$ and $\gamma_0$)  is a constant of motion:
$$
{d\xi\over dt} = 0.
\tag
$$
From the field equations and their linearization it follows that
$$
{d\over dt} \omega(\delta l,\delta\gamma_0,\delta\varphi; \tilde \delta l,\tilde\delta\gamma_0,\tilde\delta\varphi) = 0,
\tag
$$
so the pre-symplectic structure is defined independently of the value of $t$. This is
 guaranteed by the nature of its construction \refto{covps}.\footnote{*}{In fact,
a similar computation shows that
$
{d\over dt} \theta(\delta l,\delta\gamma_0,\delta\varphi) = 0.
$
 $\theta$ is conserved because of a special feature of the Einstein-Hilbert action:
it vanishes when the field equations are satisfied.}

The form $\omega$ is degenerate precisely because $\varphi$ is subject to the constraint \(constraint).  The degeneracy directions for $\omega$ are given by
$$
\delta l=0=\delta \xi,\quad 
\delta \varphi = \varphi_{,x},
\tag deg
$$
which implies, in particular,
$$
\delta\gamma_0 = \Pi_{,x}(t_0,x_0),
\tag
$$ 
so that, 
$$
\omega(\delta l, \delta\gamma_0,\delta\varphi; 0,\Pi_{,x}(t_0,x_0),\varphi_{,x}) = 0.
\tag
$$
The vector field on  $\Gamma$ defined by \(deg) generates a 1-parameter pre-symplectic group of transformations on $\Gamma$:
$$
\varphi(t,x) \to \varphi(t,x+\lambda),\quad l\to l,\quad \xi\to\xi.
\tag trans
$$

We can interpret the pre-symplectic structure on $\Gamma$ as follows. Consider the unconstrained space $\tilde\Gamma=\{(l,\gamma_0,q,p,a_n,a_n^*),\ n=\pm 1,\pm2,\ldots\}$. Define a non-degenerate 2-form $\tilde\omega$ on this space using \(symplectic), but without imposing ${\cal C}=0={\cal \delta} C$.  It is straightforward to verify that $\tilde\omega$ is conserved by virtue of the field equations and their linearization, so that this symplectic form is $t$-independent. From $\tilde\omega$ we can read off a canonical chart for $\tilde\Gamma$; the canonical pairs are $(\phi,P_\phi)$ and $(\xi, l)$. Since $l>0$ it is convenient to define new canonical variables $(Q,P)\in {\bf R}^2$ by
$$
Q =  \ln l,\quad P = - l\xi.
\tag QP
$$
We extend the one parameter group \(trans) and its infinitesimal generator \(deg) to $\tilde\Gamma$ in the obvious way.
The pre-symplectic space of solutions $(\Gamma,\omega)$ can then be re-constructed from $(\tilde\Gamma,\tilde\omega)$  by (i) imposing the constraint,
$$
{\cal C} = \int_0^{2\pi} P_\phi\phi_{,x}=0,
\tag phi_const
$$
thus defining $\Gamma\subset\tilde\Gamma$,  and (ii) pulling back the symplectic 2-form $\tilde\omega$ to $\Gamma$, thus defining $\omega$.

To summarize, the space of solutions to the vacuum Einstein equations for the Gowdy metrics \(Gowdy_metric) consists of a ``point particle'' degree of freedom with canonically conjugate variables $(Q,P)$,  and a field degree of freedom described by $\varphi$, or equivalently the canonical variables $(\phi,P_\phi)$,  or equivalently the variables $(q,p,a_n,a_n^*)\ n = \pm1, \pm2,\ldots$, subject to the constraint \(constraint), or \(phi_const).  This characterization of the vacuum Gowdy spacetimes is essentially the same as obtained from the Hamiltonian methods of \refto{CGT1996, Pierri, Corichi2002}. The only significant differences  are as follows. First, by working directly with the space of solutions to the field equations, the metric variable $\gamma$ can be defined in terms of $\varphi$  using an integral over time (and space) rather than the purely spatial integral that is used in the Hamiltonian formalism. Second, the ``point particle'' degrees of freedom $(Q,P)$ have been defined as constants of the motion and have been clearly disentangled from the time variable. There is no ``q-number'' aspect to the time, such as arises in \refto{Pierri, Corichi2002}.\footnote*{These authors rescale the time variable with a dynamical variable constructed from the point particle degree of freedom so as to simplify the form of the field equations for the field $\varphi$. However, this complicates the dynamical behavior of the point particle degrees of freedom. By working with constants of motion, these complications are avoided.}

Dynamical evolution from $t=t_1$ to $t=t_2$ can be viewed as a pre-symplectic map ${\cal T}_{t_1,t_2}\colon \Gamma\to\Gamma$ (see, \eg \refto{CGT1999c}) .  The map takes  a given solution $(l,\gamma_0,\varphi)$ of the Einstein equations  to a solution $( l, \gamma_0,\hat \varphi)$ whose Cauchy data at $t=t_1$ are the Cauchy data for $(l,\gamma_0,\varphi)$  at $t=t_2$. In particular:
$$
\eqalign{
\hat\varphi(t_1)&=\varphi(t_2)\cr
 t_1 \hat\varphi_{,t}(t_1)&=t_2\varphi_{,t}(t_2).}
\tag CT
$$ 
Since the general solution to the field equations is known explicitly, it is straightforward to construct  $(\hat l,\hat\gamma_0,\hat \varphi) = {\cal T}_{t_1,t_2}(l,\gamma_0,\varphi)$. We have
$$
\hat \gamma_0 = \gamma_0,\quad \hat l = l \quad\Longleftrightarrow\quad \hat Q = Q,\quad \hat P=P,
\tag dyn1
$$
$$
\hat \varphi(t,x)={1\over \sqrt{2\pi l}}(\hat q + \hat p \ln t)
+{1\over2\sqrt{2l}} \sum_{n=-\infty\atop n\neq 0}^\infty\left( \hat a_n  H_0(|n|t)e^{in x} + \hat a^*_n  H^*_0(|n|t)e^{-in x}\right),
\tag dyn2
$$
where
$$
\hat q = q + p\ln ({t_2\over t_1}),\quad \hat p = p,
\tag dyn3
$$
and
$$
\hat a_n = \alpha_n a_n + \beta_n a^*_{-n},
\tag dyn4
$$
with
$$
\alpha_n = {i\pi\over 4} |n|\Big[t_1H_1^*(|n| t_1)H_0(|n|t_2)
- t_2H_0^*(|n| t_1)H_1(|n|t_2)\Big]
\tag dyn5
$$
and
$$
\beta_n = {i\pi\over 4} |n|\Big[t_1H_1^*(|n| t_1)H_0^*(|n|t_2)
- t_2H_0^*(|n| t_1)H_1^*(|n|t_2)\Big].
\tag dyn6
$$
The same equations, \(dyn1)--\(dyn6) define a symplectic map  $ \tilde{\cal T}\colon \tilde\Gamma\to \tilde \Gamma$.

\taghead{3.}
\head{3. Polarized Gowdy model: quantum theory}

The quantization of the polarized Gowdy model used in  \refto{Berger, Husain1987, Pierri, Corichi2002} can be understood  in the present formulation of the model as follows.  Define a Hilbert space ${\cal H}$ by
$$
{\cal H} = L^2({\bf R}^2) \otimes {\cal F},
\tag
$$
where ${\cal F}$ is the symmetric Fock space built from the Hilbert space of square-summable complex sequences,
$
 \zeta_n, n=\pm1, \pm2, \ldots, 
$
$$
\sum_{n\neq 0} |\zeta_n|^2 < \infty.
\tag
$$
Any $\Psi\in\cal F$ can be represented as an infinite sequence of complex sequences\footnote*{Here $\psi_0$ is the ``vacuum amplitude'', $\psi_{m_1}$ is the amplitude for ``1-particle with momentum $m_1$'', \etc  For convenience, we represent the entire sequence $\{\psi_k,\ k = \pm 1, \pm2, \ldots\}$ simply by the symbol $\psi_k$.}
$$
\Psi = (\psi_0,\psi_{m_1},\psi_{m_1m_2},\ldots,\psi_{m_1\cdots m_k},\ldots),
\tag Psi
$$
where $\psi_0\in {\bf C}$,
$$
\psi_{m_1\cdots m_k} = \psi_{(m_1\cdots m_k)},
\tag symm
$$
and
$$
|\psi_0|^2 + \sum_{k=1}^\infty \, \sum_{m_1\cdots m_k\neq 0}|\psi_{m_1\cdots m_k}|^2 < \infty.
\tag norm
$$
The canonical pairs $(Q,P)$ and $(q,p)$  are represented as identity operators on $\cal F$ and are represented on (a dense domain in) $L^2({\bf R}^2)$ exactly as one would canonical coordinates and momenta for a particle moving in two dimensions, \eg
$$
\psi=\psi(x,y)\in L^2({\bf R}^2),\quad q\psi = x\psi,\quad p\psi = {1\over i}\partial_x\psi,\quad Q\psi=y\psi,
\quad
P\psi = {1\over i}\partial_y\psi.
\tag qprep
$$
(Other equivalent representations are, of course, possible.) 
 The remaining degrees of freedom in the field $\varphi$ are represented as identity operators on $L^2({\bf R}^2)$ and are represented on the Fock space $\cal F$ as follows.  
Using the representation \(Psi) for $\Psi\in {\cal F}$, we define annihilation and creation operators for each $l\neq 0$ by
$$
\eqalignno{
a_l\Psi &= (\psi_l, \sqrt{2} \psi_{lm_1},\sqrt{3}\psi_{lm_1m_2},\ldots) &(ann)\cr
a_l^*\Psi &= 
(0, \psi_0\delta_{m_1l},\sqrt{2}\delta_{l(m_1}\psi_{m_2)},\sqrt{3}\delta_{l(m_1}\psi_{m_2 m_3)},\ldots).&(cre) }
$$
These operators (on their common domain) satisfy
$$
a_n^* = (a_n)^\dagger,\quad [a_n,a_m^*] = \delta_{nm} I.
\tag a_alg
$$
The quantum field $\varphi$ is defined as an operator-valued distribution on ${\bf R}^+\times {\bf S}^1$ using the operator representation of $(q,p,a_n,a_n^*)$ in the expansion \(psi_soln). 
This quantization just described satisfies the prescription 
$$
\{{\rm Poisson\ Bracket}\}\leftrightarrow {1\over i}\ \{{\rm commutator}\}
$$ 
for the canonical coordinates on $\tilde \Gamma$, where the Poisson algebra  is defined by the symplectic form $\tilde\omega$. 

The representation of $(a_n,a_n^*)$ on $\cal F$ just described can be viewed as the outcome of a general procedure, in which the Fock space representation is defined once the appropriate ``one particle Hilbert space'' is extracted from the space of solutions of the equation \(phi_eq) (see, \eg \refto{Wald1994}).   Let us sketch this construction since some of its ingredients will be useful in what follows. Modulo the zero frequency modes $(q,p)$, the space of solutions to 
\(phi_eq)  can be identified with the space ${\cal S}$ of rapidly decreasing sequences of complex numbers $\rho\equiv\{\rho_n\}$, $n=\pm1,\pm2,\ldots$.  The symplectic form $\tilde\omega$ can be pulled back to give a symplectic form $\Omega$ on $\cal S$.
Introduce a scalar product $\mu\colon {\cal S}\times {\cal S}\to {\bf R}$ by
$$
\mu(\rho,\sigma) = \half \sum_{n\neq0} (\rho_n^* \sigma_n + \sigma_n^* \rho_n).
\tag mu
$$
Following the prescription found in reference \refto{Wald1994}, this scalar product can be used to define a Hilbert space $\bf H$ of square summable sequences of complex numbers (denoted as before) such that the Hilbert space scalar product is given  by\footnote*{The bilinear forms $\mu$ and $\Omega$ on $\cal S$ are extended to $\bf H$ by complex linearity and continuity.}
$$
(\rho,\sigma) = \mu(\rho,\sigma) - {i\over 2}\Omega(\rho,\sigma).
\tag IP
$$
$\bf H$ is the one particle Hilbert space out of which the symmetric Fock space $\cal F$ is constructed via tensor products and direct sums, as usual.  

The relation between this way of describing the Fock representation and that of references \refto{Pierri, Corichi2002} is as follows. In references \refto{Pierri, Corichi2002} the one particle Hilbert space is extracted from the space of solutions $\cal S$ of \(phi_eq) by defining a suitable complex structure $J\colon \cal S\to \cal S$.  This complex structure can be obtained from the scalar product $\mu$ and symplectic structure $\Omega$ by ``raising an index'' on $\Omega$ using $\mu$ so that
$$
\Omega(\rho,\sigma) = 2\mu(\rho,J\sigma).
\tag J
$$


As noted above, this quantization is based upon the canonical commutation relations associated to the symplectic space $(\tilde\Gamma,\tilde\omega)$, which is the space of solutions to all field equations except for the constraint \(constraint).  The usual strategy for imposing the constraint is  to represent $\cal C$ as a suitable operator on $\cal H$ and then to define the Hilbert space of physical states, $\Hphys$, as the kernel of $\cal C$.  To do this we proceed in a slightly roundabout fashion which is technically more convenient and which provides a simple example of a general strategy we shall use again when discussing the Hamiltonian.  
Essentially, we shall define ${\cal C}$ as the generator of the unitary transformation that implements the classical transformation \(trans).  The physical Hilbert space is then defined as the set of vectors which are invariant under the unitary group. 

The transformation \(trans), extended in the obvious way from $\Gamma$ to $\tilde\Gamma$, corresponds to the symplectic transformation
$$
(q,p)\to (q,p), \quad (Q,P)\to (Q,P),\quad (a_n,a_n^*)\to (e^{in\lambda}a_n,\ e^{-in\lambda}a_n^*).
\tag xtrans
$$
It is easy to check that this 1-parameter group is strongly continuous\footnote*{The transformation group $\rho\to \rho(\lambda)$ is strongly continuous if and only if $\lim_{\lambda\to \lambda_0} ||\rho(\lambda)-\rho(\lambda_0)||^2 = 0,\ \forall \lambda_0\in {\bf R}$.} in $\lambda$ relative to the norm 
$$
||\rho||^2 =\mu(\rho,\rho)
\tag munorm
$$ 
defined on {\bf H}.  It then follows from the results of \refto{Helfer1999} that the symplectic group \(xtrans) is implemented in the sense that there exists a (strongly) continuous group of unitary transformations $U(\lambda)\colon{\cal F}\to {\cal F}$ -- uniquely determined up to a phase for each $\lambda$ -- such that
$$
U^\dagger(\lambda)a_nU(\lambda) = e^{in\lambda} a_n.
\tag
$$
In fact, since the Fock representation is the GNS representation of a state canonically built from $\mu$ \refto{Wald1994}, unitary implementability follows directly from the observation that the symplectic transformation preserves the scalar product $\mu$ \refto{Bratteli1981}.  Choosing the phases so that the Fock vacuum state is invariant under the unitary group, we have, using the representation \(Psi),
$$
U(\lambda)\Psi = (\psi_0,e^{-im_1\lambda}\psi_{m_1},e^{-i(m_1+m_2)\lambda}\psi_{m_1m_2},\ldots,
e^{-i(m_1+m_2+\cdots +m_k)\lambda}\psi_{m_1m_2\cdots m_k},\ldots ).
\tag
$$
The quantized constraint (again denoted $\cal C$) is defined as 
$$
{\cal C} = i{dU(\lambda)\over d\lambda}\Bigg|_{\lambda=0}.
\tag
$$
 Using the representation \(Psi), $\cal C$ is given (on an appropriate dense domain) by
$$
{\cal C} \Psi = (0, m_1\psi_{m_1},(m_1+m_2) \psi_{m_1m_2},\ldots,
(m_1+m_2+\cdots +m_k)\psi_{m_1m_2\cdots m_k},\ldots  ).
\tag
$$ 
We extend the definition of $U(\lambda)$ and $\cal C$ to $\cal H$ by defining these operators to be the identity on the $L^2(R{\bf }^2)$ factor.
We define the physical space ${\cal H}_{phys}\subset {\cal H}$ as the set of vectors invariant under the unitary group $U(\lambda)$.
$$
{\cal H}_{phys} = \{\Psi\in {\cal H}|U(\lambda)\Psi = \Psi,\ \forall\, \lambda\in {\bf R}\}.
\tag
$$
It follows that $\Psi\in {\cal H}_{phys}$ if and only if its Fock components  \(Psi) satisfy
$$
\eqalign{
\psi_{m_1} &= 0\cr
\psi_{m_1m_2}&=\delta(m_1+m_2)\chi_{m_1},\cr
&\cdot\atop{\cdot\atop\cdot}\cr
\psi_{m_1m_2\cdots m_k}&=\delta(m_1+m_2+\cdots+m_k)\chi_{m_1m_2\cdots m_{k-1}},\cr
&\cdot\atop{\cdot\atop\cdot}\cr
}\tag phystate
$$
where the sequences $\{\psi_{m_1}, \psi_{m_1m_2}, \psi_{m_1m_2\cdots m_k}\}$ are each square-summable.
This is formally equivalent to the usual definition of ${\cal H}_{phys}$ in which $\cal C$ is defined by normal-ordering and the physical states are annihilated by $\cal C$.

\taghead{4.}
\head{4. Non-unitary dynamics on ${\cal H}_{phys}$}

The principal finding of \refto{Corichi2002} is that the Gowdy model time evolution is not unitarily implemented on the auxiliary Hilbert space $\cal H$. In our formulation of the model this result arises as follows. 

Time evolution is defined in the Heisenberg picture by the (pre-)symplectic transformation \(dyn1)--\(dyn6). Time evolution is unitarily implementable on $\cal H$ if and only if  there exists a unitary transformation $U=U(t_1,t_2)\colon{\cal H}\to {\cal H}$ such that
$$
U^\dagger QU = Q,\quad U^\dagger P U = P,
\tag unitary1
$$
$$
U^\dagger q U =  q + p\ln ({t_2\over t_1}),\quad U^\dagger p U = p,
\tag unitary2
$$
$$
U^\dagger a_n U = \alpha_n a_n + \beta_n a^\dagger_{-n},\quad 
U^\dagger a_n^\dagger U = \alpha_n^* a_n^\dagger + \beta_n^* a_{-n}.
\tag unitary3
$$
There is no obstacle to satisfying \(unitary1) and \(unitary2), but \(unitary3) is possible if and only if 
the sequence $\{\beta_n\}$ is square-summable \refto{Shale1962} (also see \refto{Wald1994} and references therein). From the large-argument asymptotic expansions of the Hankel functions appearing in $\beta_n$ it follows that  $\beta_n$ is {\it not} square-summable, 
$$
|\beta_n|^2 = {(t_2 - t_1)^2\over 4 t_1t_2} + {\cal O}({1\over n^2}) \Longrightarrow \sum_{n\neq 0} |\beta_n|^2\to \infty,
\tag beta2
$$
so $U$, as defined above, cannot exist.  

Strictly speaking, this does not establish that the time evolution fails to be unitarily implemented in the Gowdy model because the Hilbert space $\cal H$ is merely an auxiliary device used to construct the physical Hilbert space ${\cal H}_{phys}$.  Time evolution need only be unitarily implemented on ${\cal H}_{phys}$; {\it a priori} it is possible that  a unitary $U\colon {\cal H}_{phys}\to {\cal H}_{phys}$ exists, but has no appropriate extension to all of $\cal H$.  As it turns out, this scenario does not occur and time evolution cannot be unitarily implemented on ${\cal H}_{phys}$ either. So this loophole in the argument for non-unitarity given in \refto{Corichi2002} can be closed.  We demonstrate this as follows.

We first note that the product  of  any vector $\chi \in L^2({\bf R}^2)$ and  the ``vacuum state'' in $\cal F$, 
$$
\Psi_0 = \chi\otimes (1,0,0,0,\ldots),
\tag Psi0
$$
is a ``physical state'', that is,
$$
U(\lambda)\Psi_0 = \Psi_0,
\tag UPsi0
$$
so that $\Psi_0\in {\cal H}_{phys}$. Next we note that  $a_{-k}a_k$, $k\neq0$, defines a  linear operator on (a dense domain in) ${\cal H}_{phys}$. We define the time evolution of this operator to be that induced by the Bogoliubov transformation \(dyn4)--\(dyn6):
$$
a_{-k}a_k \to (\alpha_{-k} a_{-k} + \beta_{-k}a^\dagger_{k})
(\alpha_{k} a_{k} + \beta_{k}a^\dagger_{-k}).
\tag Taa
$$
It is easy to check that the right hand side of  \(Taa) defines a linear operator on (a dense domain in) ${\cal H}_{phys}.$
We now show that the square-summability condition on $\{\beta_n\}$ is again necessary for the transformation \(Taa) to be  unitarily implementable on ${\cal H}_{phys}$. To this end, we suppose that there is a family of unitary transformations 
$$
U(t_1,t_2)\colon {\cal H}_{phys}\to {\cal H}_{phys}
\tag
$$
that is continuous for all $t_1,t_2>0$ and satisfies
$$
U(t,t) = I,\quad U(t_1,t_2)U(t_2,t_3) = U(t_1,t_3),
\tag evolop
$$
$$
U^\dagger(t_1,t_2) a_{-k}a_k U(t_1,t_2) = (\alpha_{-k} a_{-k} + \beta_{-k}a^\dagger_{k})
(\alpha_{k} a_{k} + \beta_{k}a^\dagger_{-k}) .
\tag implement
$$
The condition \(implement) implies that
$$
(\alpha_{-k} a_{-k} + \beta_{-k}a^\dagger_{k})
(\alpha_{k} a_{k} + \beta_{k}a^\dagger_{-k}) U^\dagger(t_1,t_2)\Psi_0 = 0.
\tag uaa
$$
From \(UPsi0) and \(phystate) we have that $U^\dagger \Psi_0$ takes the form
$$
U^\dagger(t_1,t_2)\Psi_0 = \chi\otimes (\psi_0,0,\delta(m_1+m_2)\psi_{m_1},\delta(m_1+m_2 +m_3)\psi_{m_1m_2},\ldots),
\tag 
$$
for square-summable $\psi_{m_1}$, $\psi_{m_1m_2}$, $\ldots$.
The vacuum component of  \(uaa) implies
$$
\psi_k = -{\beta_k\over\alpha_k}\psi_0.
\tag psik
$$
From \(psik) and \(norm), a necessary condition for $U^\dagger(t_1,t_2)\Psi_0$ -- and hence $U(t_1,t_2)$  -- to be defined is square summability of $\{\psi_k\}$. Because the sequence $\{\alpha_k\}$ is bounded from below away from zero and also bounded from above, square-summability of $\{\psi_k\}$ is equivalent to square summability of $\{\beta_k\}$, assuming that $\psi_0\neq 0$.  Because $U(t_1,t_2)$ is continuous in each argument,  it follows from \(evolop) that $\psi_0\neq 0$. Therefore, unitary implementability on ${\cal H}_{phys}$ requires square summability of $\{\beta_k\}$, which does not hold, \(beta2).

\taghead{5.}
\head{5. Self-adjointness of canonical field operators}

Here is is shown that, despite the failure of time evolution to be unitarily implemented on $\cal H$ or ${\cal H}_{phys}$, the canonical coordinates and momenta for the field $\varphi$ are well-defined, self-adjoint operators for all $t>0$ and admit the usual probability interpretation.  

The canonical field operators $(\phi_{},P_{\phi})$ associated to a time $t$ are formally defined as distributions on ${\bf S}^1$ via (\cf \(psi_soln) and  \(can_field))
$$
\eqalignno{
\phi_{}(x) &= {1\over \sqrt{2\pi}}(q + p \ln t)
+{1\over2\sqrt{2 }} \sum_{n=-\infty\atop n\neq 0}^\infty\left( H_0(|n|t)e^{in x}a_n   + H^*_0(|n|t)e^{-in x}a^\dagger_n  \right),
&(ca_phi)\cr
P_{\phi}(x)&= {1\over \sqrt{2\pi}} p
-{t\over2\sqrt{2 }} \sum_{n=-\infty\atop n\neq 0}^\infty|n| \left( H_1(|n|t)e^{in x}a_n   + H^*_1(|n|t)e^{-in x}a^\dagger_n  \right).
&(can_P)}
$$
These distributions formally define canonical field operators $(\phi_{}(f), P_{\phi}(g))$ associated to any smooth real-valued functions $f$ and $g$ on ${\bf S}^1$:
$$
\phi_{}(f) =f_0(q + p \ln t) 
+{\sqrt{\pi}\over2} \sum_{n=-\infty\atop n\neq 0}^\infty\left( H_0(|n|t)f_{-n}a_n   + H^*_0(|n|t) f_{n}a^\dagger_n  \right),
\tag phi
$$
$$
P_{\phi}(g)= g_0 p
-{t\sqrt{\pi}\over2} \sum_{n=-\infty\atop n\neq 0}^\infty|n| \left( H_1(|n|t)g_{-n} a_n   + H^*_1(|n|t)g_{n} a^\dagger_n  \right),
\tag P_phi
$$
where
$$
f  = {1\over\sqrt{2\pi}}\sum_{n=-\infty}^\infty f_n e^{inx},\quad g = {1\over\sqrt{2\pi}}\sum_{n=-\infty}^\infty g_n e^{inx}.
\tag
$$
Note that the sequences $\{f_n\}$ and $\{g_n\}$ are rapidly decreasing. 

To make these formal definitions precise we consider the dense subspace
$$
{\cal H}_0 = {\cal D} \times {\cal F}_0\subset {\cal H},
$$
where ${\cal D}\subset L^2({\bf R}^2)$ is a common, invariant dense domain for the operators $(q,p, Q,P)$ (\eg in the representation  \(qprep) $\cal D$ could be chosen to be smooth functions with compact support) and ${\cal F}_0\subset {\cal F}$ is the dense set consisting of vectors with a finite number of non-zero components when  represented according to \(Psi) (\ie Fock states with ``a finite number of particles'').   It follows that ${\cal H}_0$ is a dense, invariant domain for the canonical field operators. Moreover, it is easily checked that these operators are Hermitian (\ie symmetric) on ${\cal H}_0$.  

Because $q$ and $p$ are already defined as self-adjoint operators which commute with $a_n$ and $a_n^\dagger$, in order to show that $\phi_{}(f)$ and $P_{\phi}(g)$ are self-adjoint it is sufficient to restrict attention to the field operators modulo the constant modes, which amounts to using test functions $f$ and $g$  in \(phi) and \(P_phi) with $f_0=0=g_0$.
 We then proceed by  using  the approach described in \S X.7 of \refto{Reed1975}:  prove that ${\cal F}_0$ is an analytic domain for $\phi_{}(f)$ and $P_{\phi}(g)$ (without the constant modes).   Nelson's analytic vector theorem then implies essential self-adjointness of the canonical field operators and hence a unique self-adjoint extension.  The details follow.

Begin with a Fock state with exactly $N$ ``particles'':
$$
\Psi = (0,0,\ldots,\psi_{m_1\cdots m_N},0,0,\ldots)\in {\cal F}_0.
$$
With $t$ fixed but arbitrary, define
$$
a(fH_0) = \sum_{n\neq 0} f_{-n} H_0(|n|t)a_n,\quad a^\dagger(fH_0) = \sum_{n\neq 0} f_{n} H^*_k(|n|t)a_n^\dagger
$$
It is straightforward to verify the inequalities
$$
||a(fH_0)^p\Psi|| \leq \sqrt{(N+p)!\over N!} ||fH_0||^p ||\psi||,\quad ||a^\dagger(fH_0)^p\Psi|| \leq \sqrt{(N+p)!\over N!} ||fH_0||^p ||\psi||,
\tag aest
$$
where
$$
||fH_0||^2 = \sum_{n\neq 0} |f_n H_0(|n|t)|^2,\quad ||\psi||^2 = \sum_{m_1\cdots m_n\neq0}|\psi_{m_1\cdots m_n}|^2.
\tag norms
$$
The first sum in \(norms) converges provided $\{f_n\}$ is square-summable, which it is since $f\colon {\bf S}^1\to {\bf R}$ is smooth. The second sum in \(norms) converges since $\{\psi_{m_1\cdots m_n}\}$ must be square-summable if $\Psi$ is to be in $\cal H$. 
The inequalities \(aest) guarantee that
$$
||\phi_{}(f)^p\Psi||\leq \pi^{p\over 2} \sqrt{(N+p)!\over N!}||fH_0||\, ||\psi||,
\tag phiest
$$
By definition, $\Psi$ is an analytic vector for $\phi_{}(f)$ when
$$
\sum_{l=0}^\infty ||\phi_{}(f)^l\Psi||\, {s^l\over l!} <\infty\quad \forall\ s.
\tag anal
$$
From \(phiest) it follows that
$$
||\phi_{}(f)^l\Psi||\, {s^l\over l!}\leq {s^l\over l!} \pi^{l\over 2} \sqrt{(N+l)!\over N!}||fH_0||\, ||\psi||,
\tag
$$
from which it follows that \(anal) is satisfied. This result easily generalizes to superpositions of vectors with different (but finite) numbers of ``particles'' and thence to all of ${\cal F}_0$.  

The preceding paragraph shows that, with respect to $\phi_{}(f)$, ${\cal H}_0$ is a dense, invariant, analytic subspace of $\cal H$. Therefore  $\phi_{}(f)$ is essentially self-adjoint on this domain and has a unique self-adjoint extension \refto{Reed1975}.  In a similar fashion one can show that $P_{\phi}(g)$ can be defined, for all time $t>0$, as a self-adjoint operator for any smooth function $g$.  The only change needed in the argument given above is the replacement
$$
 H_0(|n|t) f_n \to  t |n| |H_1(|n|t) g_n.
\tag
$$
Indeed,
as long as the spatial smearing functions are smooth,  all spacetime derivatives of $\varphi$ can, for each $t>0$, be defined as self-adjoint operator-valued distributions on ${\bf S}^1$.  

The spectrum of each of the canonical field operators and their derivatives is the whole real line, for all $t>0$. This can be seen, for example,  by comparing expressions such as \(phi) and \(P_phi) with their counterparts for a free  field on a flat spacetime in inertial coordinates, which have continuous spectra for all choices of the test functions. The Hankel functions appearing in \(phi) and \(P_phi) can then  be viewed as simply redefining the test functions. 
 
The spectral theorem  \refto{Reed1980} guarantees that for each self-adjoint operator $A$ on a Hilbert space $\cal H$ there is a unique-projection-valued measure $\sigma(\Omega)$ associated to any measurable set $\Omega\subset {\bf R}$ such that
$$
A = \int_{\bf R} \lambda\, d\sigma(\lambda),
\tag
$$
Given a state represented by the unit vector $\Psi\in {\cal H}$, the probability ${\cal P}_A(\Omega)$ that the observable represented by $A$ is found to take the value in the set $\Omega\subset {\bf R}$ is given by the expectation value
$$
{\cal P}_A(\Omega) = (\Psi,\sigma(\Omega)\Psi).
\tag
$$
Temporarily ignoring the constraint and working on $\cal H$, this result can be applied to the (spatially smeared) canonical field operators  at each time $t>0$, whence the usual probability interpretation can be implemented.  Note, in particular, that probabilities always add up to unity because of the general property
$$
\sigma({\bf R}) = I,
\tag
$$
where $I$ is the identity on $\cal H$.  

Of course, $\cal H$ is not the physical Hilbert space ${\cal H}_{phys}$, nor do the canonical field operators represent observables since they are not linear operators on ${\cal H}_{phys}$. The point is, however, that the failure of unitary implementability of dynamics on $\cal H$ does not destroy the physical viability of the field operators (ignoring the constraint).  Taking account of the constraint is technically more complicated, but does not alter this conclusion.  Self-adjoint operators on ${\cal H}_{phys}$ can be defined by composing polynomials in the (spatially smeared) canonical field operators at each time $t>0$ with the projection operator into ${\cal H}_{phys}$ \refto{Pierri}.   From the spectral theorem, these operators represent observables with the usual probability interpretation, despite the lack of unitary dynamics on ${\cal H}_{phys}$.  Another example of  a self-adjoint  operator on ${\cal H}_{phys}$ is  provided by the time-dependent Hamiltonian, which we shall study next.

\taghead{6.}
\head{6. Self-adjointness of the Hamiltonian(s)}

Here it is shown that  each element of the 1-parameter family of Hamiltonians for the classical Gowdy model can be promoted to a  self-adjoint operator on $\cal H$ and/or on ${\cal H}_{phys}$. At first sight, this result seems to contradict the failure of time evolution to be unitarily implemented. There is no contradiction, however. The usual link between unitary transformations and self-adjoint generators (Stone's theorem \refto{Reed1980}) is, precisely, that every continuous 1-parameter unitary {\it group} has a  self-adjoint generator, and {\it vice versa}.  Because the classical Hamiltonian depends explicitly upon the time $t$, the dynamical evolution \(dyn1)--\(dyn6) does not form a 1-parameter group and Stone's theorem does not apply.  

The classical Hamiltonian for the polarized Gowdy model can be viewed as a 1-parameter family of functions $H(t)\colon \tilde \Gamma\to {\bf R}$, which can be expressed in any of the following equivalent forms:
$$
\eqalign{
H(t) &= l E(t)\cr
&=\int_0^{2\pi} dx\, \half \left({1\over t}P_\phi^2 + t\phi^{\prime2}\right)\cr
&=  {1\over 2t}p^2 + \sum_{n\neq 0}\left(
\half A_n(t) a_n a_{-n} + B_n(t) a_n^* a_n + \half A_n^*(t) a^*_n a^*_{-n}\right),}
\tag
$$
where
$$
A_n(t) = {\pi t\over 4} n^2\left[(H_0(|n|t))^2 + (H_1(|n|t))^2\right]
\tag
$$
$$
B_n(t) = {\pi t\over 4} n^2\left[|H_0(|n|t)|^2 + |H_1(|n|t)|^2\right].
\tag
$$
The functions $H(t)$ are generators of the transformation \(dyn1)--\(dyn6) in the following sense. The infinitesimal form of the transformation \(dyn1)--\(dyn6) at time $t$ is given by\footnote\dag{Note that the variation does not commute with the time derivative since the form of the canonical transformation (like its generating function) depends upon time.}
$$
\delta l = 0 = \delta \gamma_0 \Longleftrightarrow \delta Q = 0 = \delta P,
\tag
$$
$$
\delta \varphi = \varphi_{,t},\quad \delta (t\varphi_{,t}) = {\partial\over\partial t}(t\varphi_{,t}).
\tag
$$
We have, for any tangent vector $(\delta l,\delta \gamma_0, \delta\varphi)$, 
$$
\tilde\omega(\delta l,\delta \gamma_0, \delta\varphi;  0,0,\varphi_{,t}) = \delta H(t).
\tag
$$

To define $H(t)$ as a family of operators on $\cal H$, we proceed as we did when defining the constraint operator $\cal C$. For each fixed time $t=\tau$ we compute the 1-parameter symplectic group on $(\tilde\Gamma,\tilde\omega)$ generated by $H(\tau)$.\footnote*{We emphasize that this group is {\it not} the set of time evolution canonical transformations.} As we shall see, this one parameter symplectic group can be implemented as a continuous unitary group on $\cal H$. The infinitesimal generator -- the Hamiltonian --  can then be defined via Stone's theorem. Finally, it is easy to check that this unitary group preserves ${\cal H}_{phys}$, so that the Hamiltonians thus defined are  self-adjoint operators on ${\cal H}_{phys}$.  Here are the details.

The 1-parameter group generated by $H(\tau)$ can be viewed as a transformation on $\tilde\Gamma$
$$
(l,\gamma_0,q,p,a_k,a_k^*)\longrightarrow (l(s),\gamma_0(s),q(s),p(s),a_k(s),a_k^*(s))
\tag
$$
defined by (`` $\dot{}$ '' $\equiv {d\over ds}$)
$$
\tilde\omega(\delta l,\delta \gamma_0,\delta\varphi,\dot l,\dot\gamma_0,\dot\varphi) = \delta H(\tau),
\tag
$$
so that
$$
\dot Q(s) = \dot P(s)  = \dot p(s) = 0,\quad \dot q(s) = {p(s)\over \tau}
\tag
$$
$$
\eqalignno{
\dot a_m(s) &= -i B_m(\tau) a_m(s) - i A_m^*(\tau) a_{-m}^*(s)
&(adot1)\cr
\dot a_{m}^*(s) &= i B_m(\tau) a_{m}^*(s)+ i A_m(\tau) a_{-m}(s).
&(adot2)}
$$
The point particle degrees of freedom  $(Q,P)$ are group invariants. The zero-frequency field modes $(q,p)$ transform as do the coordinate and momentum of a free particle with mass $\tau$ under time evolution.  These transformations are certainly  implementable as a continuous one-parameter unitary group. The transformation of the non-zero frequency field modes  remains to be considered.  The solution of \(adot1)--\(adot2) is given by
$$
\eqalign{
a_m(s) &=\left[ \cos(|m|s)-i {B_m(\tau)\over|m|}\sin(|m|s )\right] a_m(0)
- {i A^*_m(\tau)\over|m|}\sin(|m| s) a_{-m}^*(0),\cr a_m^*(s) &= (a_m(s))^*.}
\tag stransform
$$
Using the theory of unitary implementability \refto{Shale1962, Wald1994} of symplectic transformations, the transformation \(stransform) is, for each $s$,  unitarily implementable if and only if 
$$
\sum_{n\neq 0}  \left|{A_m(\tau)\over m}\sin(|m| s)\right|^2 < \infty.
\tag sum
$$
It is straightforward to check that
$$
\left|A_m(\tau)\right|^2 = {1\over 4 \tau^2} + {\cal O}({1\over n^2}),
\tag Aapprox
$$
so that \(sum) is satisfied.  

Furthermore, the transformation \(stransform) can be implemented as a continuous, unitary, 1-parameter group if it is strongly continuous in the norm \(munorm) \refto{Helfer1999}. 
Strong continuity means
$$
\lim_{s\to s_0} ||a(s) - a(s_0)||^2 = \lim_{s\to s_0} \sum_{n\neq0} |a_n(s) - a_n(s_0)|^2 = 0,
\tag sc
$$
which is easily verified as follows. The Bogoliubov coefficients in \(stransform) are bounded  so that there is an $n$ and $s$-independent constant such that
$$
|a_n(s)| \leq (const.)(|a_n(s_0)| + |a_{-n}(s_0)|).
\tag
$$
Therefore
$$
\eqalign{
|a_n(s) - a_n(s_0)|^2 &\leq (|a_n(s)| + |a_n(s_0)|)^2 \cr
&\leq (const.) [|a_n(s_0)|^2 + |a_{-n}(s_0)|^2 + |a_{-n}(s_0)||a_n(s_0)|].}
\tag
$$
The right hand side of this inequality defines a square summable sequence of real numbers, thanks to the square-summability of $a_n(s_0)$. (Square summability of the first two terms is obvious, the last follows from the Schwarz inequality.) By the Weierstrass M-test \refto{Apostol1974}  this guarantees that 
$\sum |a_n(s) - a_n(s_0)|^2$ converges uniformly  for all $s$. Uniform convergence guarantees that $\sum |a_n(s) - a_n(s_0)|^2$ converges to a continuous function of $s$, implying \(sc).

From all these considerations, \(stransform) is implementable as a continuous unitary group $U(s)\colon {\cal H}\to {\cal H}$, 
$$
U^\dagger(s) a_m U(s) = \left[ \cos(|m|s)-i {B_m(\tau)\over|m|}\sin(|m|s )\right] a_m
- {i A^*_m(\tau)\over|m|}\sin(|m| s) a_{-m}^\dagger,
\tag uofs 
$$
from which the Hamiltonian $H(\tau)$ is uniquely defined, up to an additive multiple of the identity,  as the infinitesimal generator.

The continuous unitary group $U(s)\colon {\cal H}\to {\cal H}$ acts on
the physical Hilbert space.\footnote*{Formally, this follows from the fact that ${\cal C}$ and $H(\tau)$ commute.   But without a precise characterization of the domain of $H(\tau)$ it is hard to  conclude anything from this formal result.} 
We can then say that the Hamiltonian $H(\tau)$ represents an observable (although it is not known precisely what is the domain of $H(\tau)$ in ${\cal H}_{phys}$).
To see that $U(s)$ acts on ${\cal H}_{phys}$, we first note that the image of the state $\Psi_0$ defined in \(Psi0) is a physical state, $U(s)\Psi_0\in {\cal H}_{phys}$. Indeed, writing \(uofs) as
$$
U^\dagger(s) a_m U(s) = \alpha_m(s) a_m + \beta_m(s) a_{-m}^\dagger.
\tag uau
$$
a straightforward computation shows that 
$$
U(s) \Psi_0 = N(s)\exp\left\{-i {p^2\over 2\tau}s -\half \sum_{n\neq 0} \gamma_n(s) a_n^\dagger a_{-n}^\dagger\right\}\Psi_0,
\tag u0
$$
where
$$
\gamma_n(s) = {\beta_n(-s)\over \alpha_n(-s)}
$$
and $N(s)$ is fixed (up to a phase) by normalization.
Evidently, the action of $U(s)$ on  the Fock vacuum is given by ``pair creation'' with each pair having zero total momentum, thus yielding a state in ${\cal H}_{phys}$.   Using \(uau) and \(u0) it is straightforward to compute the action of $U(s)$ on a vector obtained as the image of any polynomial in the creation operators applied to $\Psi_0$.  Since these states span $\cal H$ (as $\chi$ varies over a basis for $L^2({\bf R}^2)$), this defines $U(s)$. It is then easy to see that states satisfying \(phystate) are mapped  by $U(s)$ into states satisfying \(phystate).  For example,
$$
\eqalign{
U(s) a_n^\dagger a_{-n}^\dagger\Psi_0= \Big[&\alpha_n^{*2}(-s) a_{-n}^\dagger a_n^\dagger\cr
 &+ \alpha_n^*(-s)\beta_n^*(-s)(a_n^\dagger a_n + a_{-n}a_{-n}^\dagger)
+ \beta_n^{*2}(-s) a_{-n}a_n\Big]U(s)\Psi_0,}
\tag
$$
from which it is clear that
$$
U(s) a_n^\dagger a_{-n}^\dagger\Psi_0 \in {\cal H}_{phys}.
$$

\taghead{7.}
\head{7. Remarks on the physical viability of the model}

In the quantum mechanical description of systems with a finite number of degrees of freedom, lack of unitary dynamics is normally associated with a failure of the probability interpretation of the model. The absence of unitary time evolution in the Gowdy model is an ultraviolet effect of the same sort as observed in \refto{Helfer1996, CGT1999c}; it has no analog in the quantum mechanics of a system with a finite number of degrees of freedom.  Indeed, we have seen that, despite the lack of unitary dynamics, the probability interpretation of the quantum Gowdy model  appears to be intact in the following sense. The basic dynamical variables $(Q,P)$ and $(\varphi,\varphi_{,t})$ (with the latter smeared with smooth functions of $x$) are self-adjoint operators on $\cal H$ for all $t>0$.  Observables (self-adjoint operators on  ${\cal H}_{phys}$) -- besides functions of $(Q,P)$ --    can be built from the field variables $(\varphi,\varphi_{,t})_t$ by projection.  The spectral theorem then guarantees that the set of possible outcomes of a measurement of the observables  have probabilities which add up to unity for all $t>0$.  Remarkably, even more complicated observables such as the Hamiltonians $H(t)$ for the model can, for each $t>0$, be defined as  self-adjoint operators and given a consistent probability interpretation.  Similar remarks can be made about the systems considered in \refto{Helfer1996, CGT1999c}. There, the basic linear fields can, for all time, be defined as self-adjoint operators, so that the usual probability interpretation is available for them,  despite the fact that the time evolution is not unitarily implemented.  

In \refto{Jacobson1991} it is argued that non-unitary Schr\"odinger picture evolution in quantum gravity leads to difficulties with causality and locality. It is also pointed out there that these difficulties may be absent in a formulation of dynamics in the Heisenberg picture.  Of course, in quantum mechanics the mathematically distinct Schr\"odinger and Heisenberg pictures are physically equivalent. But this is  precisely because of the unitary implementation of dynamics in either picture.  In the type of situation being discussed in this article, dynamical evolution is defined in the Heisenberg picture by the field equations, and the Schr\"odinger picture description of dynamics is unavailable.  In this way the Gowdy model, as well as the models appearing in \refto{Helfer1996, CGT1999c},  appear to evade the unacceptable behavior discussed in  \refto{Jacobson1991}.

For these reasons, it appears that the lack of unitary dynamics need not render the quantum Gowdy model  physically unacceptable. Still, an important question left open by these considerations is the status of fundamental geometrical quantities in the Gowdy model, \eg the metric and curvature .  Some of the features of the metric and curvature operators have been studied in \refto{Berger, Husain1987, Pierri}, but a more thorough investigation is warranted.  To give a flavor of the issues involved, let us consider the metric components in \(Gowdy_metric) from the point of view of the quantum theory.  Evidently, the exponentials of the quantum fields $\varphi$ and $\gamma$ are required to define the quantum metric.  Of course, it is too much to ask that the quantum metric components be defined by self-adjoint operators pointwise; some smeared version is required.  Since $Q$ and $\phi(f)$ (defined in \(phi)) are self-adjoint, we can define a smeared self-adjoint $\varphi$ via
$$
\varphi(f) = e^{-\half Q}\phi(f),
\tag
$$
which can be exponentiated to define a self-adjoint, smeared metric component. To define the quantum $g_{yy}$ and $g_{zz}$ metric components  in an arbitrarily small neighborhood of any point one need only choose a sufficiently well-localized smearing function $f$.   The definition of the metric function $\gamma$ (needed for $g_{tt}$ and $g_{xx}$) is more problematic. The field $\gamma$ is formally defined in terms of $(Q,P,\phi,P_\phi)$ via eqs. \(gamma_soln), \(can_field), \(xi), and \(QP). It is not clear that smearing $\gamma$ with a smooth function of $x$ and, \eg normal-ordering of the creation and annihilation operators, will be adequate to render the expressions involving $\cal E$ and $\Pi$ well-defined.  That question aside, it is not at all clear that the variable $\gamma_0$, defined in terms of $(Q,P,\phi,P_\phi)$ via \(xi) is well-defined. For example, the integral appearing in the last term of \(xi) is the generator for the following 1-parameter group of  canonical transformations
$$
\phi \to \phi(\alpha)=  e^\alpha \phi,\quad P_\phi\to P_\phi(\alpha) = e^{-\alpha}P_\phi.
\tag
$$
It is straightforward to verify that this transformation group is not unitarily implemented on $\cal H$, precluding the existence of a self-adjoint generator. Evidently if $\gamma$  can be defined at all it will be through some sort of regularization procedure yet to be constructed (\cf \refto{Pierri}). It would seem that the need for a physically tenable definition of $\gamma$ in the quantum theory is the salient difficulty with the Gowdy model, not the failure of unitarity. 

Supposing one could not satisfactorily resolve the question of how to define $\gamma$ in the quantization of the Gowdy spacetimes under consideration, one might respond by searching for a different representation of the canonical commutation relations (CCR) than that used thus far, one which allows for a well-defined metric operator.  Moreover, it is possible that other representations of the CCR will also allow for unitarily implemented time evolution.\footnote*{This was suggested in reference \refto{Corichi2002}.    Some of Berger's work \refto{Berger} can be interpreted as considering alternate representations of the CCR.}  However, the need for a search for an optimal choice of representation can be eliminated if one is willing to define the quantum Gowdy model using the algebraic quantum field theory (AQFT) formalism, such as been advocated by Wald \refto{Wald1994}, who was motivated by issues arising in quantum field theory in curved spacetime. In this formalism one, in effect, uses all representations of the CCR at once. Questions of unitary implementability become moot in the algebraic approach since they are tied to a particular representation, which is physically irrelevant from the point of view of AQFT.  Indeed, it is sufficient to demonstrate that time evolution defines an automorphism of the $C^*$ algebraic structure used to define the theory, in which case the observables appearing in the $C^*$ algebra will be defined as self-adjoint operators for all time in any representation.  An application of AQFT to the Gowdy model  will have to take account of the non-linear nature of the space of solutions to the field equations -- the constraint in particular, and the need to give probability distributions for the metric, curvature, \etc This certainly seems feasible and will be pursued elsewhere. 

\bigskip\noindent
{\bf Acknowledgments}

Thanks to Karel Kucha\v r and Chris Beetle for helpful discussions. This investigation was supported in part by grant PHY-0070867 from the National Science Foundation. \bigskip

\references

\refis{Kuchar1971}{K. V. Kucha\v r, \prd 4, 955, 
1971.} 

\refis{Misner1973}{C. Misner, \prd 8, 3271, 1973.}

\refis{Berger}{B. Berger, \ann 83, 458, 1974; 
\prd 11, 2770, 1975; \ann 156, 155, 1984.}

\refis{Husain1987}{V. Husain, \cqg 4, 1587, 1987.}

\refis{Husain1989}{V. Husain and L. Smolin, \npb 327, 205, 1989.}

\refis{Neville1993}{D.~E.~Neville, \cqg 10, 2223, 1993.}

\refis{CGT1996}{J. D. Romano and C. G. Torre, 
\prd 53, 5634, 1996.}

\refis{Ashtekar1996b}{A. Ashtekar and M. Pierri,
\jmp 37, 6250, 1996.}

\refis{Korotkin1998}{D. Korotkin and H. Samtleben, \prl  80, 14, 1998.}

\refis{Pierri}{M. Pierri, {\it Int. J. Mod. Phys. D}  {\bf 11}, 135 (2002);
gr-qc/0201013 (2002).}

\refis{Corichi2002}{A. Corichi, J. Cortez and H. Quevedo, gr-qc/0204053 (2002).}

\refis{Helfer1996}{A. Helfer, \journal Class. Quantum Grav., 
13, L129, 1996.}

 \refis{CGT1999c}{C. G. Torre and M. Varadarajan, \journal 
 Class. Quantum Grav., 16, 2651, 1999.}

\refis{Jacobson1991}{T. Jacobson, in {\it 
Conceptual 
Problems 
of Quantum Gravity}, edited by A. Ashtekar and J. Stachel, 
(Birkh\"auser, 
Boston 1991).}

\refis{Hawking1973}{S. Hawking and G. Ellis, {\it The Large 
Scale Structure of Space-Time}, (Cambridge University 
Press, Cambridge 1973).}

\refis{Gowdy1974}{R.~Gowdy, \ann 83, 203, 1974.}

\refis{Helfer1999}{A. Helfer, preprint hep-th/9908011.}

\refis{covps}{See, for example, C. Crnkovic and E. Witten in {\it 300 
Years of 
Gravitation}, 
edited by S. Hawking and W. Israel (Cambridge University 
Press, 
Cambridge 1987); 
J. Lee and R. Wald, \jmp 31, 725, 1990;
A. Ashtekar, L. Bombelli, and O. 
Reula, in {\it Mechanics, 
Analysis and Geometry : 200 Years After Lagrange}, edited 
by M.
Francaviglia ( North-Holland, New York 1991); 
G. Barnich, M. Henneaux, C. 
Schomblond, \prd 44, R939, 1991.}

\refis{Chrusciel1990}{P. Chrusciel, J. Isenberg and V. Moncrief, \cqg 7, 1671, 1990.}

\refis{Gowdy1974}{R.~Gowdy, \ann 83, 203, 1974.}

\refis{Beetle1998}{C. Beetle,  \journal Adv. Theor. Math. Phys., 2, 471, 1998.}

\refis{CGT2002}{M. E. Fels and C. G. Torre, \cqg 19, 641, 2002.}

\refis{Wald1994}{R. Wald, {\it Quantum Field Theory in 
Curved Spacetime and Black Hole Thermodynamics}, University 
of Chicago Press, 1994.}

\refis{Bratteli1981}{O. Bratteli and D. Robinson, {\it Operator Algebras and Quantum Statistical Mechanics}, (Springer-Verlag, New York 1981).}

\refis{Shale1962}{D. Shale, {\it Trans. Am. Math. Soc.}, {\bf 103}, 149 (1962).}

\refis{Reed1975}{M. Reed and B. Simon, {\sl Methods of Modern Mathematical Physics, Vol. II}, (Academic Press, New York, 1975).}

\refis{Reed1980}{M. Reed and B. Simon, {\sl Methods of Modern Mathematical Physics, Vol. I}, (Academic Press, New York, 1980).}

\refis{Apostol1974}{T. Apostol, {\it Mathematical Analysis}, (Addison-Wesley, Reading, 1974).}

\refis{Segal1963}{I. Segal, in {\it Mathematical Problems of Relativistic Physics, Vol. II}, Mark Kac, Ed. (American Mathematical Society, Providence 1963).}

\endreferences

\endit